\documentclass[preprint,aps,showpacs,prb,floatfix]{revtex4}

\usepackage{graphicx}
\usepackage{dcolumn}
\usepackage{bm}

\begin{document}

\preprint{}

\title{Enhancement of entanglement in one-dimensional disordered systems}
\author{Rom\'an L\'opez-Sandoval$^1$
 and Martin E. Garcia$^2$}
\affiliation{$^1$Instituto Potosino de Investigaci\'on Cient\'{\i}fica
y Tecnol\'ogica, Camino a la presa San Jos\'e 2055, 78216
San Luis Potos\'{\i}, Mexico}
\affiliation{$^2$ Theoretische Physik, FB 18, Universit\"at Kassel and 
 Center for Interdisciplinary Nanostructure Science and Technology 
(CINSaT),
Heinrich-Plett-Str.40, 34132 Kassel, Germany}

\date{\today}

\begin{abstract}
The pairwise quantum entanglement of sites in disordered electronic 
one-dimensional systems (rings) is studied. We focus on 
the effect of diagonal and off diagonal disorder on the concurrence
$C_{ij}$ between electrons on neighbor and non neighbor sites
$i,j$ as a function of band filling.  In the case of
diagonal disorder, increasing the degree of disorder leads to a
decrease of the concurrence with respect to the ordered case. However,
off-diagonal disorder produces a surprisingly strong enhancement of
entanglement. This remarkable effect occurs near half filling, where
the concurrence becomes up to 15\% larger than in the ordered 
system. 
\end{abstract}
%


\pacs{03.67.Mn,71.23.An,71.23.-k}
\maketitle

\section{Introduction}
\label{sec:introd}
It has been shown that entanglement is a necessary requirement for
the realization of several algorithms used in quantum computation and
information. For example, quantum teleportation, which is a technique
for moving a quantum state between two points, requires the use of
entangled states\cite{Nielsen}. On the other hand, there are some
evidences that many problems involving strong-coupled many body
systems can be characterized and classified by using quantum
entanglement. Therefore, this can lead to new and more effective
methods for understanding the dynamical behavior of complex quantum
systems \cite{Preskill}. In particular, it has been conjectured that
entanglement can play an important role in quantum phase
transitions \cite{Osterloh,GVidal,Gu,JVidal}.

Several relations between entanglement and quantum phase transitions
have been obtained recently. Osterloh et al., have analyzed the
behavior of entanglement near the critical point of the spin 1/2 model
XY in a transverse field \cite{Osterloh}. They have found that in the
region close to a quantum phase transition the entropy of entanglement
obeys a scaling law. Although the entanglement is not an indicator of
that phase transition, they showed that there exists an intimate
connection between entanglement, scaling and universality. Moreover,
Vidal et al. \cite{GVidal}, have established a precise connection
between concepts of quantum information, condensed matter physics, and
quantum field theory by showing that the behavior of critical
entanglement in spin systems is analogous to that of entropy in
conformal field theory. In addition, Gu et al \cite{Gu} have studied the
properties of entanglement in the ground-state of an antiferromagnetic
XXZ chain. They have shown that the competition between quantum
fluctuation and ordering leads to a maximization of entanglement at
the isotropic point.

Interestingly, also in the case of electron systems a quantitative
analysis of entanglement can be performed. Recently, Zanardi et
al. studied the pairwise entanglement between nearest-neighbor sites
in itinerant spinless-fermion systems described by a periodic
tight-binding Hamiltonian\cite{Zanardi}. They calculated the
concurrence, which is a quantity which provides the same information
as the entropy of entanglement\cite{Wootters,Connor}, and showed that
the maximal concurrence between two neighboring sites is obtained at
half band filling $x=n/N_{a}=0.5$, being the $n$ the number of
electrons and $N_a$ the number of atoms (sites). The concurrence is
symmetric with respect to the half-filling point.

The impurities or disorder effects have been seen in quantum
computation as something not desired that can induce errors in gating
or in the evolution of a quantum state \cite{Montangero}.  In
addition, if the imperfection strength is increased, new phenomena
related with chaotic behavior can occur \cite{Georgeot}. On the other
hand, it is well known that  crystalline systems with impurities can
undergo  a metal-insulator transition as a function of disorder strength
\cite{Review}. Therefore, it is interesting to study   
the effect of disorder on the concurrence and if it is possible to 
find a relation between both.

In this paper, we present the first study of entanglement in
disordered electronic systems.  We analyze the effect of disorder on
the entanglement between sites in one-dimensional rings. For this
purpose, we consider diagonal and off-diagonal disorder and different
degrees of disorder.  We show that whereas diagonal disorder reduces
and suppresses entanglement for all values of the band filling $x$,
off-diagonal disorder considerably enhances the entropy of
entanglement for $x$ near, but not exactly at, half-filling.

The paper is organized as follows.  In section ~\ref{sec:theo}, we
discuss the employed model Hamiltonian and how to quantify the
entanglement.  In Sec.~\ref{sec:size}, the effect of the length of the
ring on the entanglement is studied. We analyze the effect of diagonal
and off-diagonal disorder on entanglement in Sec.~\ref{sec:diagdis}
and \ref{sec:nodiagdis}. Finally, in Sec. ~\ref{sec:conc}, we
summarize our results.

\section{Theory}
\label{sec:theo}
We consider one-dimensional disordered electronic systems and describe
them by a Anderson-Hamiltonian of the form\cite{anderson}
\begin{equation}
\label{eq:Ham}
{\hat{H}}=\sum_{i}  \varepsilon_i  \hat n_i
+ \sum_{\langle ij \rangle}  t_{ij}   \hat c^{+}_i \hat c_j .
\end{equation}
For simplicity, we consider spinless electrons. In Eq.~(\ref{eq:Ham})
$\hat c^{+}_i$ ($\hat c_i$) is the usual creation (annihilation)
operator of a spinless electron at site $i$, whereas $\hat n_i =\hat
c^{+}_i \hat c_i $ is the number operator, and $t_{ij}$ is the hopping
integral between nearest neighbor (NN) sites $i$ and $j$.
$\varepsilon_i$ is the on-site energy for atom $i$.  Diagonal disorder
is introduced by assuming that the values of $\varepsilon_i$ are
randomly distributed in the energy interval $[0,W]$ while the NN
hopping integrals remain unchanged with respect to the perfectly
ordered case, i.e., $t_{ij}=-1$. Analogously, the effect of
off-diagonal disorder (random hopping) is modeled by considering the
hopping integrals $t_{ij}$ to acquire random values in the interval
$[0,W]$ while all on-site energies are equal to zero, like in the
ordered lattice.

To calculate the entanglement of formation between pairs of sites we
resort to the concept of concurrence\cite{Wootters,Connor} and proceed
in a similar way as Zanardi and coworkers did for the case of
translationally invariant chains\cite{Zanardi}.

Basically, one can consider the ring with $N_a$ sites and $n$
electrons as being a bipartite system composed of two subsystems, $A$
and $B$, where $A$ refers to the pair of sites $i$ and $j$ and $B$ to
the rest of the ring (see Fig.~1). More precisely, $A$ contains all 4
possible electronic states in the pair of sites $i$ and $j$, i.e.,
$|1\rangle_A \equiv |0\rangle_A$, $|2\rangle_A \equiv c^{+}_i
|0\rangle_A$, $|3\rangle_A \equiv c^{+}_j |0\rangle_A$, and
$|4\rangle_A \equiv c^{+}_i c^{+}_j |0\rangle_A$. Analogously, $B$
contains the states of the rest of the ring. Thus, a particular state
$|\Psi_{AB}\rangle $ of the entire ring can be represented as a direct
product of the form
\begin{equation}
\label{eq:state}
|\Psi_{AB} \rangle=
 \sum_{\mu \nu} \alpha_{\mu \nu} |\psi_{A\mu} \rangle \otimes
 |\psi_{B\nu} \rangle,
\end{equation}
where $ |\psi_{A\mu} \rangle$ and $|\psi_{B\nu} \rangle$ are basis
states of the subsystems $A$ and $B$. Thus, the density matrix of the
whole system reads 
\begin{equation}
\label{rho}
\hat \rho = |\Psi_{AB} \rangle \langle \Psi_{AB} |. 
\end{equation} 
The reduced density matrix $\rho_A$ for the subsystem $A$ is obtained
by performing a trace on subsystem $B$ in order to integrate out its
degrees of freedom. Therefore 
\begin{equation}
\label{rhoa}
\hat \rho_A = Tr_B \left\{ |\Psi_{AB} \rangle \langle \Psi_{AB} | \right\} .
\end{equation} 
The entropy of entanglement of a bipartite system is calculated
as \cite{bennett} 
\begin{equation}
\label{eq:entropy}
 E(|\Psi_{AB}\rangle ) = 
- Tr_A \left\{ \hat \rho_A \log \hat \rho_A\right\} = 
-Tr_B \left\{ \hat \rho_B \log \hat \rho_B\right\}
\end{equation} 
and quantifies the entanglement of formation. It
has been shown by Wootters\cite{Wootters} that the entropy of
entanglement can be also characterized by the concurrence $C$, since $
E(|\Psi_{AB} \rangle )$ is a monotonous function of $C$, which is
defined by
\begin{equation}
\label{eq:concdef}
C = max\left( \lambda_1 -
\lambda_2 - \lambda_3 -\lambda_4, 0 \right).
\end{equation}
 The $\lambda_k$'s are,
in decreasing order from $k =1$ to $k=4$, the eigenvalues of the
matrix $R=\sqrt{\rho_A \, {\tilde{\rho}}_A}$, where ${\tilde{\rho}}_A
= \left( \sigma_y \otimes \sigma_y\right) \rho^*_A \left( \sigma_y
\otimes \sigma_y\right)$, where $\rho^*_A$ is the complex conjugated of
the matrix $\rho_A$ and $\sigma_y$ the ``$y$'' Pauli-matrix.

Thus, in order to calculate $C$ for disordered 1D systems described
by the Hamiltonian of Eq.~(\ref{eq:Ham}), we should first determine
$\rho_A$ by means of Eqs.~(\ref{eq:state}) and (\ref{rhoa}). As
pointed out in Ref. \onlinecite{Zanardi}, the number of electrons $n$ is
conserved due to the fact that $[{\hat{H}}, {\hat{n}}]=0$. As a
consequence, only the elements $\rho_A(11)$, $\rho_A(22)$,
$\rho_A(33)$, $\rho_A(44)$, $\rho_A(23)$ and $\rho_A(32)$ are
nonzero. All other elements vanish. This can be seen by looking at the
expansion (\ref{eq:state}). When constructing $|\Psi_{AB} \rangle
\langle \Psi_{AB} |$ and performing the trace over $B$ one obtains,
for all elements of $\rho_A$ others than the ones mentioned above,
scalar products of the type $\langle \psi_{B\nu} | \psi_{B\mu}
\rangle$, between states of $B$ having different number of particles
and, therefore, being orthogonal.
 One can take profit of this property of the reduced density matrix and 
 obtain its nonzero elements as quantum expectation values of relevant physical
 quantities. 
 For this purpose we use the whole wave function $|\Psi \rangle_{AB}$ of the 
 system and write:
 \begin{eqnarray}
\label{eq:rhoa11}
\rho_A(11) = \langle \Psi_{AB} |   (1- \hat n_i) (1- \hat n_j)|
\Psi_{AB} \rangle \\
\label{eq:rhoa44}
\rho_A(44) = \langle \Psi_{AB} |   \hat n_i \hat n_j| 
\Psi_{AB} \rangle\\
\label{eq:rhoa22}
\rho_A(22) = \langle \Psi_{AB} |   \hat n _i  (1- \hat n_j ) 
|\Psi_{AB} \rangle\\
\label{eq:rhoa33}
\rho_A(33) = \langle \Psi_{AB} |   (1- \hat n_i ) \hat  n_j| 
\Psi_{AB} \rangle\\
\label{eq:rhoa23}
\rho_A(23) = \rho^*_A(32) = \langle \Psi_{AB} | c^+_j c^{}_i |
\Psi_{AB} \rangle.
\end{eqnarray}
In Eq.~(\ref{eq:rhoa11}), the presence of the operator $(1- \hat
n_i)  (1- \hat n_j )$ automatically projects out all states of the
whole system not having 0 electrons on site $i$ and 0 electrons on
site $j$. Similarly, the operator $\hat n_i  \hat  n_j $ in
Eq.~(\ref{eq:rhoa44}) ensures that only states having 1 electron on
$i$ and 1 electron on $j$ will contribute to the element
$\rho_A(44)$. In Eqs.~(\ref{eq:rhoa22}) and (\ref{eq:rhoa33}) the
corresponding projectors to obtain $\rho_A(22)$ and $\rho_A(33)$ are
described. Finally, $\rho_A(23)$ [$\rho_A(32)$] contains only 
contributions from states having 1 electron on site $i$ ($j$) and no
electrons on site $j$ ($i$).

Thus, with the help of the above equations, one can write reduced 
density  matrix $\rho_A$ in the following
form \cite{Zanardi,Connor}
\begin{equation}
\label{eq:rhomatrix}
\rho_A= \left( \matrix{
v  & 0 & 0  & 0  \cr
0  & w & z  & 0  \cr
0  & \bar z  & u  & 0  \cr
0  & 0 & 0  & y  
}
\right),
\end{equation}
where the nonvanishing matrix elements are given by
\begin{eqnarray}
v=1- \langle \hat n_i  \rangle - \langle \hat n_j \rangle +   
\langle \hat n_i \hat n_j  \rangle,  \quad y=
\langle \hat n_i \hat n_j \rangle  \nonumber \\
z= \langle \hat c^{+}_j \hat c_i \rangle, 
\quad w = \langle \hat  n_i  \rangle - y, \quad
 u = \langle \hat  n_j   \rangle - y.
\label{eq:rhoaelements}
\end{eqnarray}
Eqs.~(\ref{eq:rhoaelements}) show how the nonzero elements of
$\rho_A$, can be calculated as average quantities of the complete
ground-state wave function.
 
Diagonalization of the matrix of Eq.~(\ref{eq:rhomatrix}) is
trivial. Using Eq.~(\ref{eq:concdef}) one obtains an analytic
expression for the concurrence between sites $i$ and $j$ of the
form\cite{Connor}
\begin{equation}
\label{eq:Conc}
C_{ij}= 2 \max \{ 0, |z|-\sqrt{vy}\; \}.
\end{equation}
Eq.~(\ref{eq:Conc}) does not contain $w$ and $u$. Note that $i$ and $j$
do not need be nearest neighbors. In the next section we present
calculations of $C_{i,i+\ell}$ with $\ell=1,3$.

For every set of parameters, the Hamiltonian can be diagonalized
numerically and the concurrence $C$ can be obtained using
Eqs.~(\ref{eq:rhoaelements}) and (\ref{eq:Conc}). The average value of
the concurrence can be then calculated by using a large number of
replicas in order to minimize numerical fluctuations.

To calculate the average concurrence of the tight-binding rings with
diagonal and off-diagonal disorder, we have implemented the following
procedure: $(i)$ we have generated a set of random parameters (on-site
energies or hopping elements) for each of the replicas in order to
model the disorder of the finite size rings, $(ii)$ we have
numerically diagonalized the Hamiltonian, given by Eq.~\ref{eq:Ham},
for obtaining its single-particle eigenvectors, $(iii)$ with the help
of these eigenvectors, we have calculated the values $v$, $y$ and $z$
necessary for the calculation of the concurrence and, finally, $(iv)$
we have obtained the average concurrence $\bar C_{ij}$
($\bar C_{12}, \bar C_{13}$, $\bar C_{14}$) as
\begin{equation}
\label{eq:averageconc}
\bar C_{1,1+\ell} = \frac{1}{N_{replicas}N_{a}} 
\sum_{replicas} \sum^{N_a}_{k} C_{k,k+\ell}.  
\end{equation}
We have performed these calculations for different ring sizes. We
found that for rings consisting of $N_a\simeq 200$ sites results
achieve convergence. Therefore, in this paper all the results for
$\bar C_{ij}$ correspond to $N_a=200$ sites, unless we explicitly
specify the ring size.  On the other hand, the number of replicas we
used was 10000.  In order to understand the influence of disorder on
the concurrence it is necessary to study first the behavior of the
concurrence in ordered systems.

\section{Ordered rings: size effects} 
\label{sec:size}
The concurrence between $C_{12}$ as a function of $n$ for a perfectly
ordered, periodic ring can be easily computed, using the theory
presented in the previous section, as a limiting case of $\bar 
C_{ij}$ for $\varepsilon_i=0 \; \forall i$ and $t_{ij}=-1$ ).
Obviously, the sums of Eq.~(\ref{eq:averageconc}) are not necessary in
this case, since they cancel with the factors in the denominator. We
have computed $C_{12}$ vs $n$ for different sizes. Results are shown
in Fig.~\ref{fig:szeff}. As an example, we show below the analytical
calculation for $N_a = 4$.

The $4 \times 4$ Hamiltonian has the eigenvalues $\epsilon_1= -2$,
$\epsilon_2= \epsilon_3= 0$, $\epsilon_4= 2$, and the
eigenvector matrix reads
\begin{equation}
\label{eq:rhomatrixa}
U= \left( \matrix{
\frac{1}{2}  & \frac{1}{\sqrt{2}} & 0  & \frac{1}{\sqrt{2}} \cr
 \frac{1}{2}& 0 & \frac{1}{\sqrt{2}}  & -\frac{1}{2}  \cr
 \frac{1}{2}& -\frac{1}{\sqrt{2}} & 0  & \frac{1}{2}  \cr
  \frac{1}{2}& 0 & -\frac{1}{\sqrt{2}}  &  -\frac{1}{2} 
}
\right). 
\end{equation}

With this information at hand it is possible to calculate the elements
of the reduced density matrix and therefore the concurrence, as
summarized in Table I.
\begin{table}
\caption{Calculation of the elements of the reduced density matrix 
$\rho_A$ and the concurrence according to equations
  \ref{eq:rhomatrix} to 
\ref{eq:rhomatrixa} , for ring consisting of 4 atoms and $n$ electrons. }
\begin{ruledtabular}
\begin{tabular}{|c|c|c|c|c|}
$n$ & 1 & 2 & 3 & 4  \\
$\langle n_1 \rangle$ & 1/4 & 3/4 & 3/4 & 1 \\
$\langle n_2 \rangle$ & 1/4 & 1/4 & 3/4 & 1 \\
$z$ & 1/4 & 1/4 & 1/4 & 0 \\
$\langle n_1 n_2\rangle$ & 0 & 1/8 & 1/2 & 1 \\
$v$ & 1/2 & 1/8 & 0 & 0 \\
$y$ & 0 & 1/8 & 1/2 & 1 \\
$C_{12}$ & 1/2 & 1/4 & 1/2 & 0 \\ 
\end{tabular}
\end{ruledtabular}
\end{table}

Clearly, the concurrence is symmetric with respect to $n=2$ (half
filling). The concurrence for $n=0$ and $n=4$ is equal to zero.  We
observe in Fig.~\ref{fig:szeff}(a) that for smaller ring ($N_a=4$, 8
and 12 sites) it is very difficult to perceive which is the 
band filling $x$ where 
$C_{12}$ will have a maximum.  In addition, the concurrence as a
function of $n$ does not have a monotonous behavior.  These size
effects can be understood with the help of the correlation functions $
z=\langle \hat c^{+}_1 \hat c_2 \rangle $, $ y=\langle \hat n_1 \hat
n_2 \rangle$, and the band filling $x$.  It is easy to show that
$z=\langle \hat c^{+}_1 \hat c_2 \rangle $ is related to the
ground-state energy $E_{\rm gs}$, i.e.  $z=\langle \hat c^{+}_1 \hat
c_2 \rangle = -E_{\rm gs}/{2N_a}$, while that $v=(x-1)^2-z^2$ and
$y=x^2-z^2$ (see Ref. 7). From these expressions, we obtain for systems
with band filling $x=1/N_a$ (only one electron in the lattice) that
$C_{12}=2z=-E_{\rm gs}/N_a$ where $y=\langle \hat n_1
\hat n_2 \rangle=0$.  This explains the value of $C_{12}$ found for
$N_a=4$ with one electron ($x=0.25$). In general, the concurrence of
periodic rings can be written as follows \cite{Zanardi} 
\begin{eqnarray}
C_{12} = 2 \max \left \{ 0,|E_{\rm gs}/2N_a| - \left( 
\left [ (x-1)^2 -(E_{\rm gs}/2N_a)^2\right] 
\left[ x^2-(E_{\rm gs}/2N_a)^2
\right] 
\right)^{1/2} 
\right \}. 
\end{eqnarray}
When we begin to increase the ring size, we notice that the finite
size effect begins to dissapear [see
Fig.~\ref{fig:szeff}(b)]. Moreover, $C_{12}$ begins to have a
monotonous behavior and we find that the maximum value of NN
concurrence occurs at half-band filling. For $N_a=64$ sites the
thermodynamic limit has been reached and the values of $C_{12}$ are
almost converged. 

\section{Disordered rings}

\subsection{Diagonal disorder}
\label{sec:diagdis}
The electronic properties of strongly disorder systems have been
investigated by a variety of analytical and numerical methods. In this
paper, we study this problem from a very different perspective, the
perspective of the concurrence. In Fig.~\ref{fig:c12diads} we show the
NN concurrence $\bar C_{12}$ for a ring with $N_a=200$ sites
where we have introduced only diagonal disorder. We observe that the
effect of this diagonal disorder is to decrease $\bar C_{12}$ in
a monotonous way.  In Fig~\ref{fig:c12diads}(a) $\bar C_{12}$ is
presented as a function of band filling for some representative values
of the disorder strength $W$. This decrease of the concurrence is
related to the localization of the wave function. When disorder is
introduced, for example on site $i$, the wave functions are localized
around this site, therefore the correlation function $z=\langle \hat
c^{+}_i \hat c_j \rangle$ between two NN sites is decreased. On the
other hand, Osterloh et al.
\cite{Osterloh},  in their study of the critical point in the XY model  
in a transversal magnetic field, showed that the behavior of the
concurrence does not have a divergence in the transition
point. However, they showed that the derivative of $C_{12}$ shows this
divergence and scaling properties. Therefore, it may be of interest to
study the derivative of $\bar C_{12}$ as a function of the disorder
strength. Unfortunately, it is known that numerical calculation of
derivatives in disorder systems as a function of $W$ displays errors
associated with the numerical division \cite{Taylor}. In order to 
visualize the behavior of the derivative of $C_{12}$ as a function of
the strength of disorder, in Fig~\ref{fig:c12diads}(b) we plot $\bar
C_{12}$ as a function of the disorder strength $W$ for some
representative values of the band filling $x$.  From the figure, we
can deduce that the derivative of $\bar C_{12}$ as a function of
disorder strength is almost zero in the range of $W \in
[0,2.5]$. After that, the derivative of $\bar C_{12}$ as a function of
$W$ is almost constant with a slight negative slope. 
A similar analysis can be realized on the concurrence as a function of
the distance between sites.  In particular, we have focused on next
nearest-neighbor concurrence $\bar C_{13}$ and third nearest-neighbor
concurrence $\bar C_{14}$.  The results are shown  as a function of
the band filling $x$ for some representative values disorder strength
$W$ [Figs. \ref{fig:c13diads}(a) and \ref{fig:c14diads}(a)] and as a
function of the disorder strength $W$ for some values of the band
filling $x$ [Figs. \ref{fig:c13diads}(b) and
\ref{fig:c14diads}(b)]. We observe that  
the values of the concurrence for all band fillings and disorder
strengths is much smaller than $\bar C_{12}$.  This implies that in
systems without disorder and systems with diagonal disorder the
concurrence is highly local and strongly decreases as a function of
the distance between sites. Notice that the maximum value of the
concurrence occurs at $W=0$ and begins to move towards band fillings
smaller than the half band 1/2  when we increase the distance
between sites and disappears in some range of band filling. This
result can be explained by using the fact that the correlation
function $z=\langle
\hat c_{i}^{+} \hat c_{j}
\rangle$ depends on the overlap between the wave function, which 
decreases when we increase the distance between sites, whereas $v$ and
$y$ fundamentally depend on the band filling. Therefore, it will have
a threshold band filling where $\sqrt{vy} > |z|$ and, consequently,
$\bar C_{ij}=0$. On the other hand, as we begin to increase the
disorder strength, the concurrence is almost insensitive to the system
band fillings.  These behaviors can be observed in figures 4(b) and
5(b), where we show the concurrence for systems with different band
fillings and $W>10$ are joined in the same curve. This result is
related to the fact that due to the strong disorder $z$, $v$ are $y$
small for all band fillings.

\subsection{Off diagonal disorder} 
\label{sec:nodiagdis} 
Although for 1D systems with diagonal or off-diagonal disorder the
wave function are localized for all disorder strengths, it is known
that the physical properties are strongly dependent on the type of
disorder.  Systems with off-diagonal disorder show an anomalous
behavior in the density of states and in the localization length
\cite{Soukoulis,Inui}, where it has been found that $E=0$ state is
localized but has an infinite localization length. In this subsection,
we show how this property affects the concurrence. 

In Fig.~\ref{fig:c12odiadis} we present results for $\bar C_{12}$ for a
ring with $N_a=200$ sites and with off-diagonal disorder $t_{ \rm NN}
\in [0,W]$. In Fig.~\ref{fig:c12odiadis}(a) $\bar C_{12}$ is showed as a
function of band filling $x$ for some representative disorder
strengths $W$ whereas in Fig.~\ref{fig:c12odiadis}(b) $\bar C_{12}$ is
presented as a function of $W$ for some representative values of
$x$. From the figures, we observe that, contrary to the diagonal
disorder, $\bar C_{12}$ value can increase for some band fillings in
comparison with the non disordered case ($W=0$) when the disorder
strength is increased.  In particular, we observe that this occurs for
band fillings larger than $x \ge 0.25$ and for off-diagonal disorder
strength $W \ge 4$. Notice that the maximum value of the concurrence
is found for band filling values in the range $ 0.4 \le x \le 0.45$
and for $W\ge 20$. Moreover, the overall form of the concurrence
curves does not change anymore for $W > 20$. It is surprising that in
the case of system with off-diagonal disorder, the NN concurrence
increases in comparison with respect to the non disordered case.
This behavior can be related to the anomaly in the density of states
found in these kind of systems. To inquire about it, we have
calculated the NN concurrence as a function of the band filling for a
small system with $N_a=16$ sites with only one different hopping
elements with respect to the others.  These results are shown in
Fig.~\ref{fig:c12oneimp}, where we present the concurrence between the
sites with the different hopping integral ($t_{12}=-2,t_{\rm NN}=-1$,
otherwise), $ C_{12}$, and the NN concurrence between sites different
from 1 and 2.  In the last case, we only show the concurrence between
sites close to the impurity. From the figure we observe that $C_{12}$
has much larger values compared to the result show in the previous
subsection.  This concurrence is almost independent of the band filling
and oscillates around 0.72. These high values of $ C_{12}$ are related
to the fact that the impurity localizes the eigenfunction $|\psi_0
\rangle $ with the lowest eigenvalue $\varepsilon_0$ between the sites
where the impurity was placed ($t_{12}=-2$), whereas the other
eigenfunctions are almost delocalized.  Thus, we have that $z= \langle
\psi_0 |c_1^{+} c_2 |\psi_0 \rangle $ has very large contribution to
the concurrence for all band fillings. On the other hand, we know that
in the case of one electron, $vy=0$. Therefore from the figure, we can
see that the quasidelocalized eigenfunctions contribute very little to
the NN concurrence when we increase the band filling.  In the case of
$C_{23}$, we observe that the localization effect turns $C_{23}$ into
a decreasing function. It is necessary to remark that the localization
effect due to one impurity only strongly modifies the NN concurrence
between sites very near to this, where the eigenfunction $ |\psi _0
\rangle$ is localized, whereas the NN concurrence of remote sites from
the impurity show a behavior more similar to the case of a periodic
ring.  Therefore, when the hopping integrals are randomly chosen in
the range $t_{\rm NN} \in [0,W]$, the number of hopping integrals
having values larger than $W/2$ is $N_a/2$. This implies that in
general there are $N_a/2$ eigenfunctions that are localized on the
bonds affected by these hopping integrals.  These $N_a/2$ bonds will
contribute with very high values to $\bar C_{\rm NN}$, as has been
shown in the case of one impurity, whereas the other $N_a/2$ bonds
will contribute with much smaller values. As we increase $W$, the
localization of the eigenfunction will be increased around these bonds
and their $\bar C_{\rm NN}$ values will be close to 1 whereas the
$\bar C_{\rm NN}$ values of the other bonds will be close to zero.
These bond-localizations explain the increase in the NN concurrences
of systems with off-diagonal disorder in comparison with the non
disordered case. In addition, this clearly shows the difference
between both types of disorder, in diagonal disorder the one particle
eigenfunctions are localized around the sites, which decreases the
total value of $\bar C_{\rm NN}$ with increasing $W$. On the other
hand, in the off-diagonal disorder case the eigenfunctions are
localized between the nearest neighbors, which increases $\bar C_{\rm
NN}$ with increasing $W$.

In order to study the way the off-diagonal disorder affects the
concurrence as a function of the distance between sites in this kind
of systems, we have calculated the concurrence as a function of the
distance. The results are shown in Figs.~\ref{fig:c13odiadis}and
\ref{fig:c14odiadis}. In these figures we observe that, as the
off-diagonal strength $W$ is increased, the maximum in the concurrence
begins to decrease, contrary to the case of the NN concurrence. This
confirms that the effect of the increase in the maxima of the NN
concurrence is due to the strong localization on the bonds. Contrary
to the case of diagonal disorder, the decrease of the maximum values
of $\bar C_{13}$ and $\bar C_{14}$ for large $W$ is relatively small
with respect to the non disordered case. We again notice that the
concurrence decreases as a function of the distance between sites,
which shows that the concurrence is a strongly local
property. Moreover in these cases, the curves of $\bar C_{13}$ and
$\bar C_{14}$ depend on the band filling. This behavior can be seen in
Figures 8(b) and 9(b), where we show that the curves for different
band fillings are completely different. Finally, for a given  band
filling $x$ and for $W>10$, the curves are almost insensitives to 
the strength of the disorder.

\section{Conclusions} 
\label{sec:conc} 
In this work, we have studied the effect of diagonal and off diagonal
disorder on 1D electronic systems described by the Anderson
Hamiltonian.  We have found that, in the case of diagonal disorder, as
we increase the disorder strength the concurrence between sites
decreases in a monotonous way. In addition, we found that the
concurrence in these systems is highly local and decays for
increasing distances between sites. On the other hand, in off diagonal
disorder case we observe that the increase in the disorder strength
results in an increase of the NN concurrence in comparison to finite
rings in the band filling interval $0.25 \le x \le 0.5$.  This
behavior of the concurrence for both kinds of disorder can be
understood with the help of the one particle wavefunctions. In the
diagonal disorder case, the one particle wave functions begin to
localize on the sites as we increase the disorder strength. Therefore,
this localization decreases the NN concurrence. On the other hand, the
increase of the off-diagonal disorder strength localizes the one
particle wavefunction on bonds between NN sites, which increases the
NN concurrence in a certain interval of the band filling.  This result 
could be useful for the study of entanglement in 3D amorphous solids. 
 Although the metal-insulator transition
occurs only in 3D systems \cite{MacKinnon}, it is known that the
off-diagonal disorder in 1D systems produces anomalous properties
\cite{Soukoulis,Inui} that should have some effect of the  
concurrence. 

Investigations of 
the concurrence as an indicator
of the insulator-metal
phase transition in 3D disordered systems are in progress.

\begin{acknowledgments}

This work has been supported by CONACyT Mexico (Grant No.\ J-41452 and
Millennium Iniative W-8001) Computer resources were provided by CNS
(IPICYT, Mexico).

\end{acknowledgments}

%


\begin{figure}
\centerline{\includegraphics[scale=0.7]{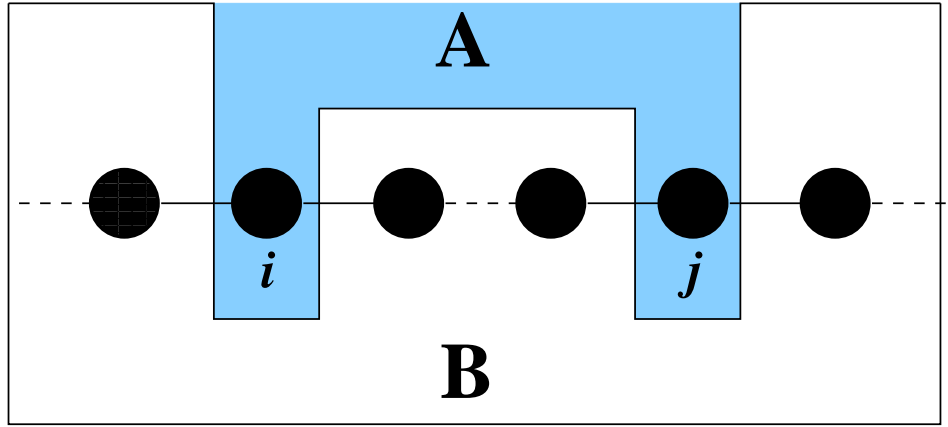}}
\caption{(Color online) 
Schematic illustration of the partition of the disordered rings into two subsystems for the  calculation of the entropy of entanglement between sites $i$ and $j$.}
\label{fig:scheme}
\end{figure}

\begin{figure}
\centerline{\includegraphics[scale=0.7]{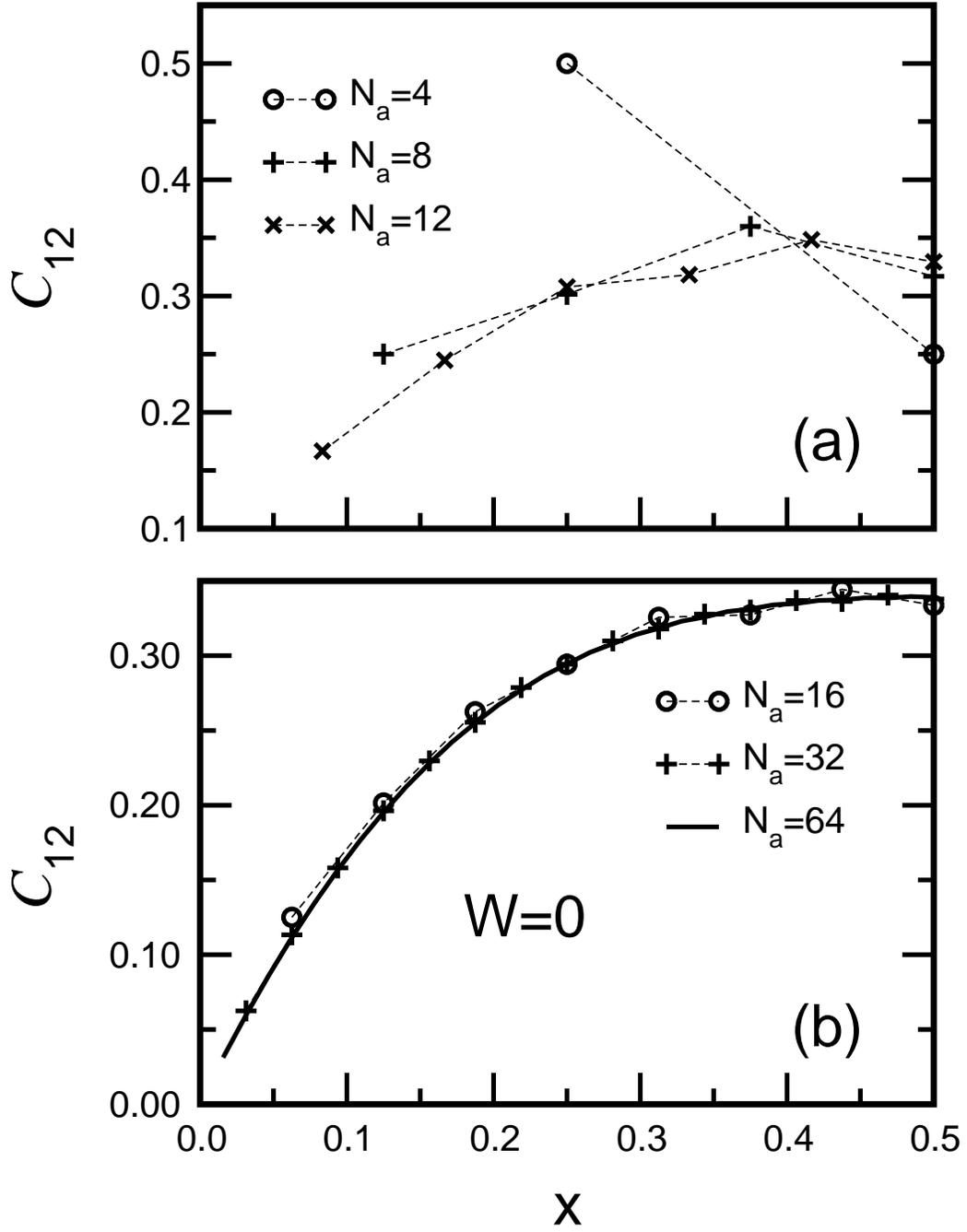}}
\caption{Nearest neighbor concurrence $C_{12}$
of small periodic ring  as a function of band
filling $x$.}
\label{fig:szeff}
\end{figure}

\begin{figure}
\centerline{\includegraphics[scale=0.6]{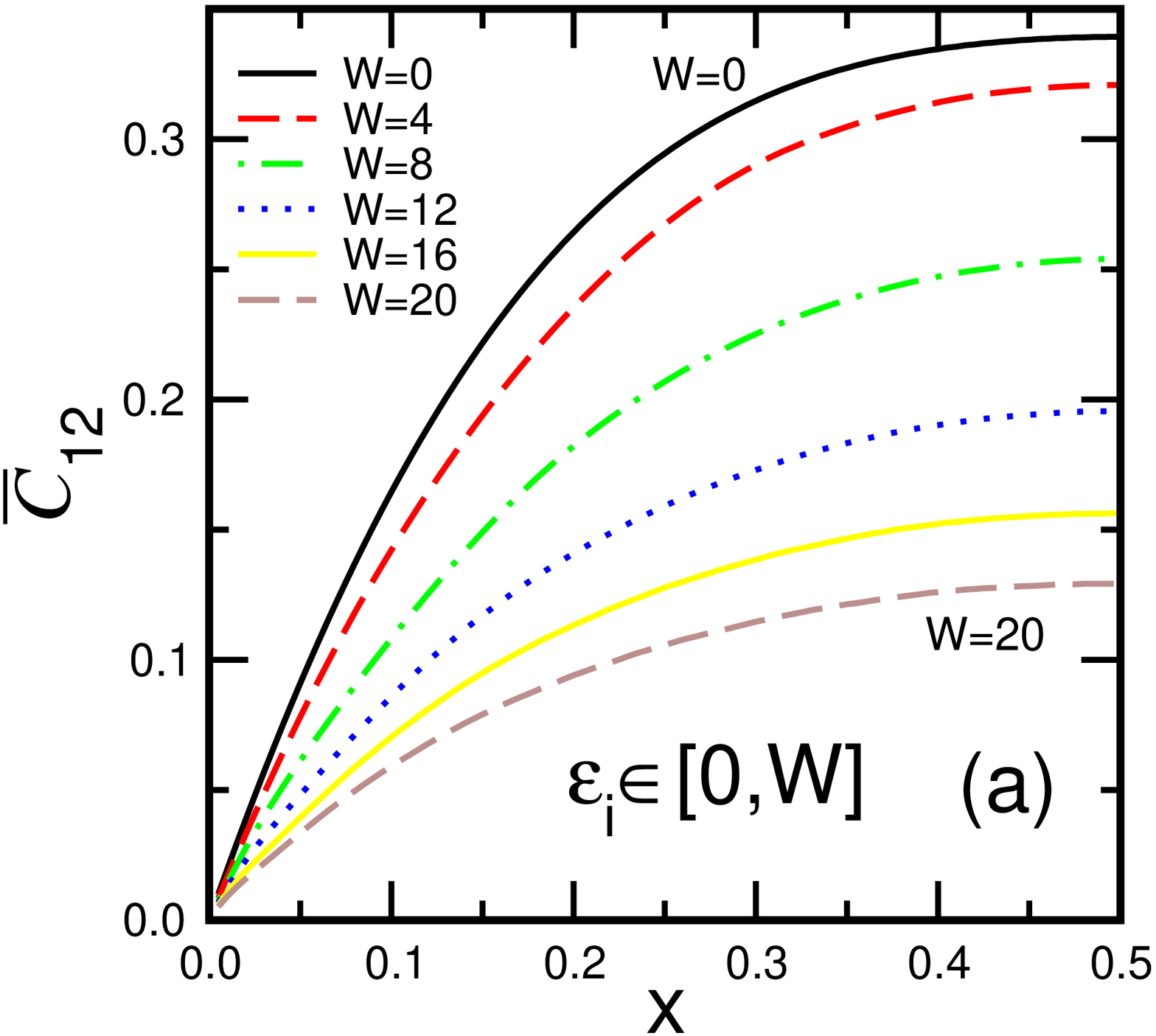}}
\centerline{\includegraphics[scale=0.6]{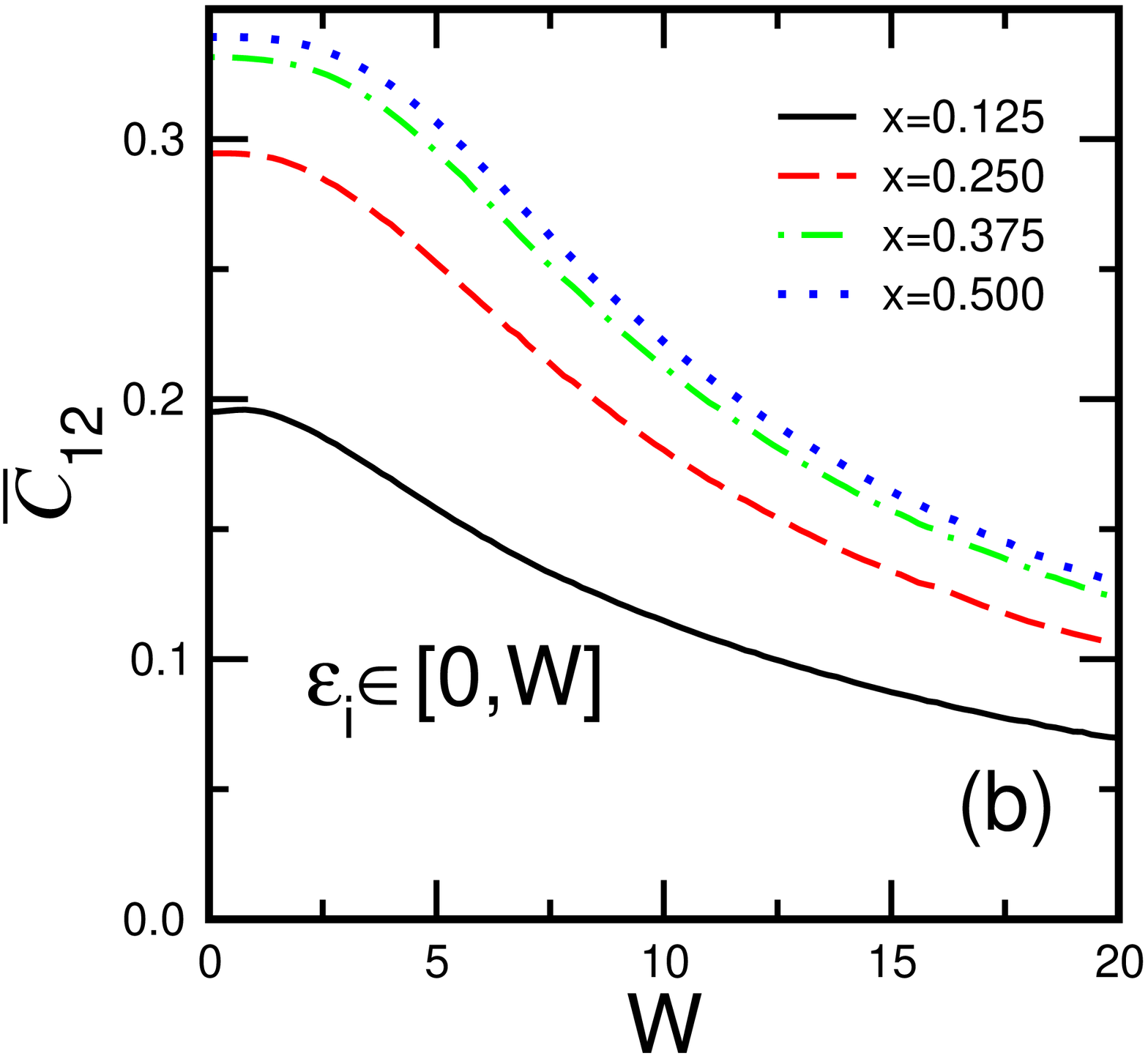}}
\caption{(Color online) 
Nearest neighbor concurrence average $\bar C_{12}$ of a
ring with $N_a=200$ sites . In (a) we present $\bar C_{12}$ as a function
of band filling $x$ for several representative values of the diagonal
disorder strength $W$. In (b) $\bar C_{12}$ is plotted as a function of
disorder strength $W$ for some band filling $x$.}
\label{fig:c12diads}
\end{figure}

\begin{figure}
\centerline{\includegraphics[scale=0.6]{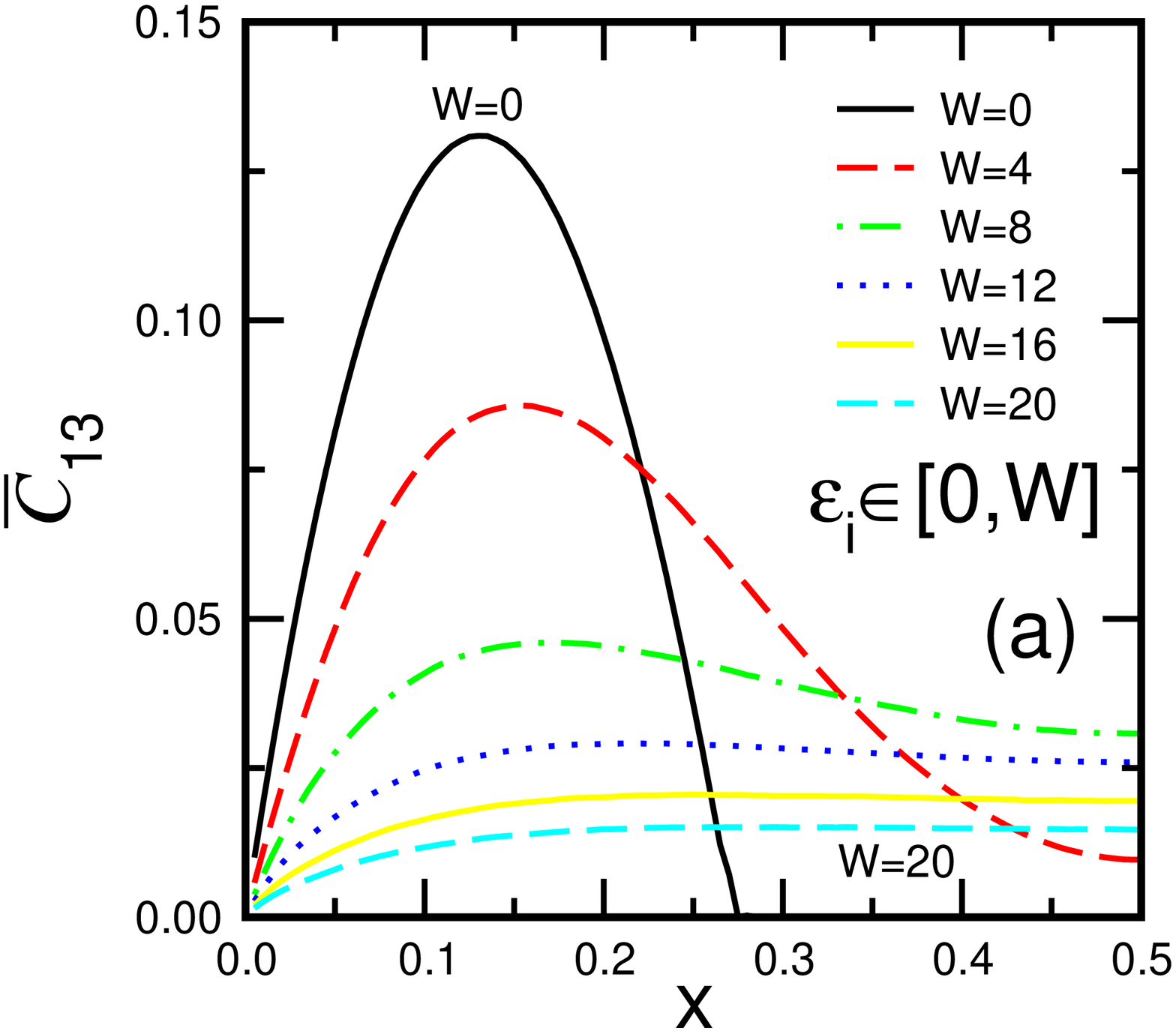}}
\centerline{\includegraphics[scale=0.6]{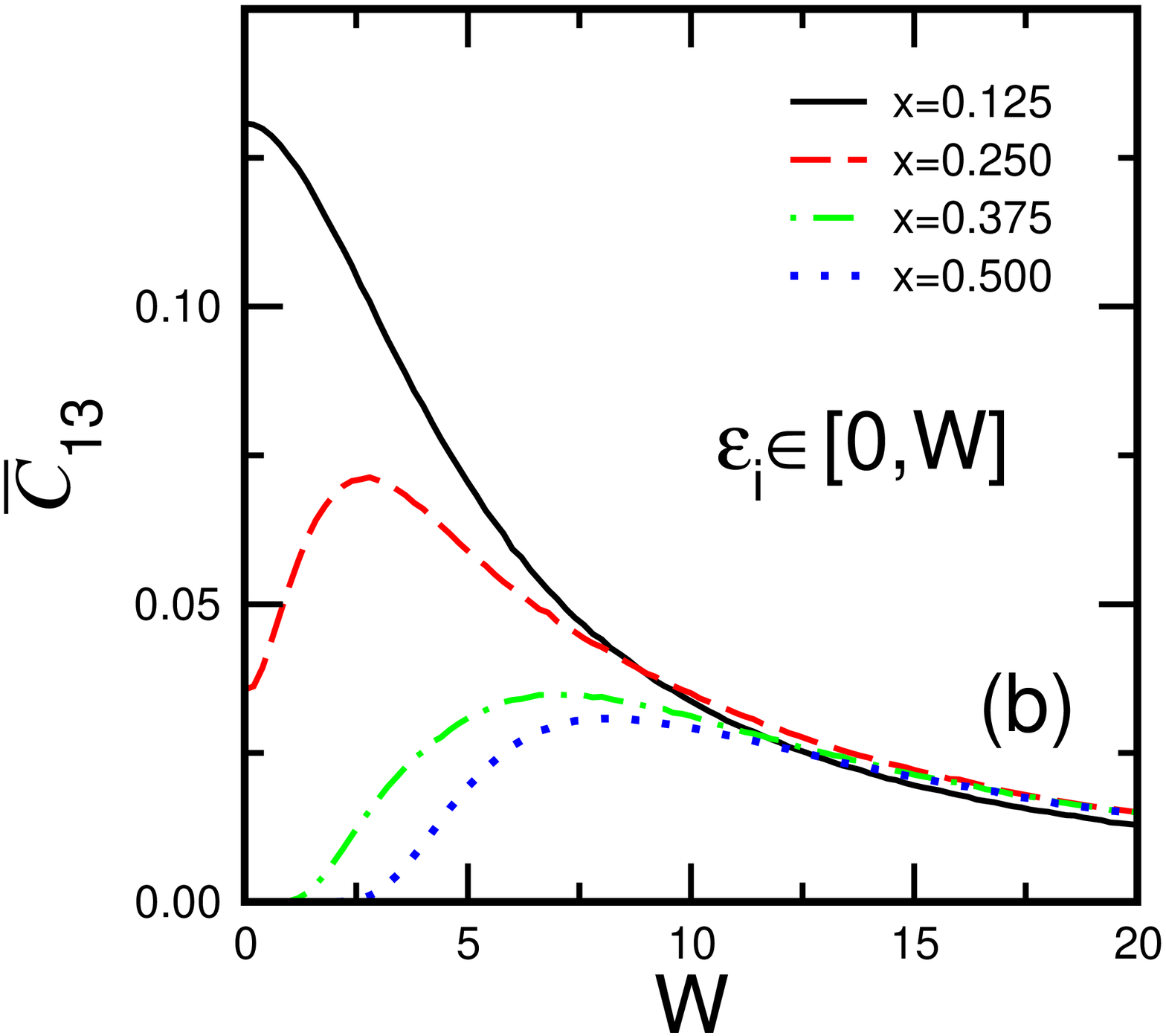} }
\caption{(Color online)  
Next nearest-neighbor average concurrence $\bar C_{13}$ of a
ring with $N_a=200$ sites as a function of band filling $x$
for some representative values of diagonal disorder
strength $W$.}
\label{fig:c13diads}
\end{figure}

\begin{figure}
\centerline{\includegraphics[scale=0.6]{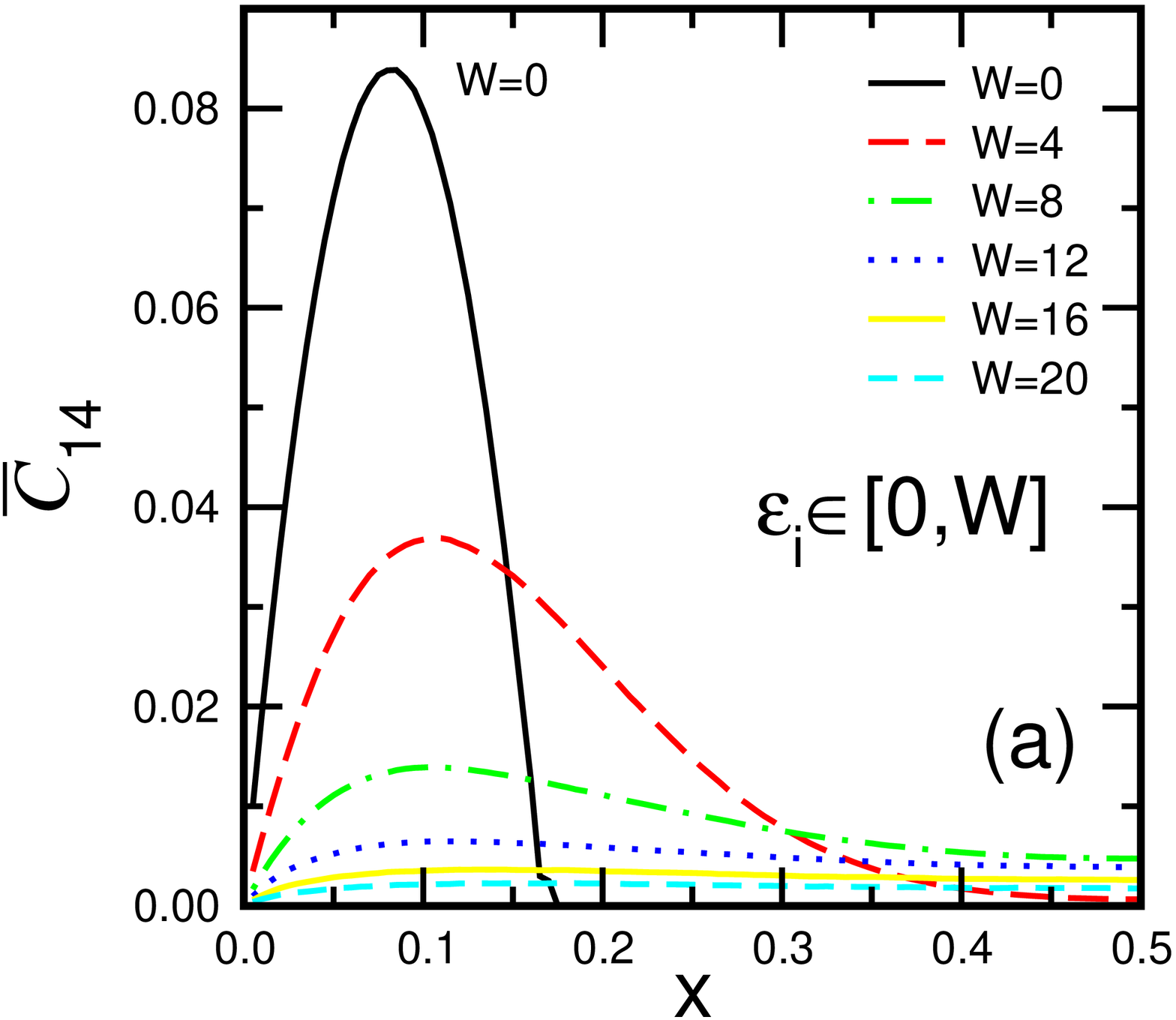}}
\centerline{\includegraphics[scale=0.6]{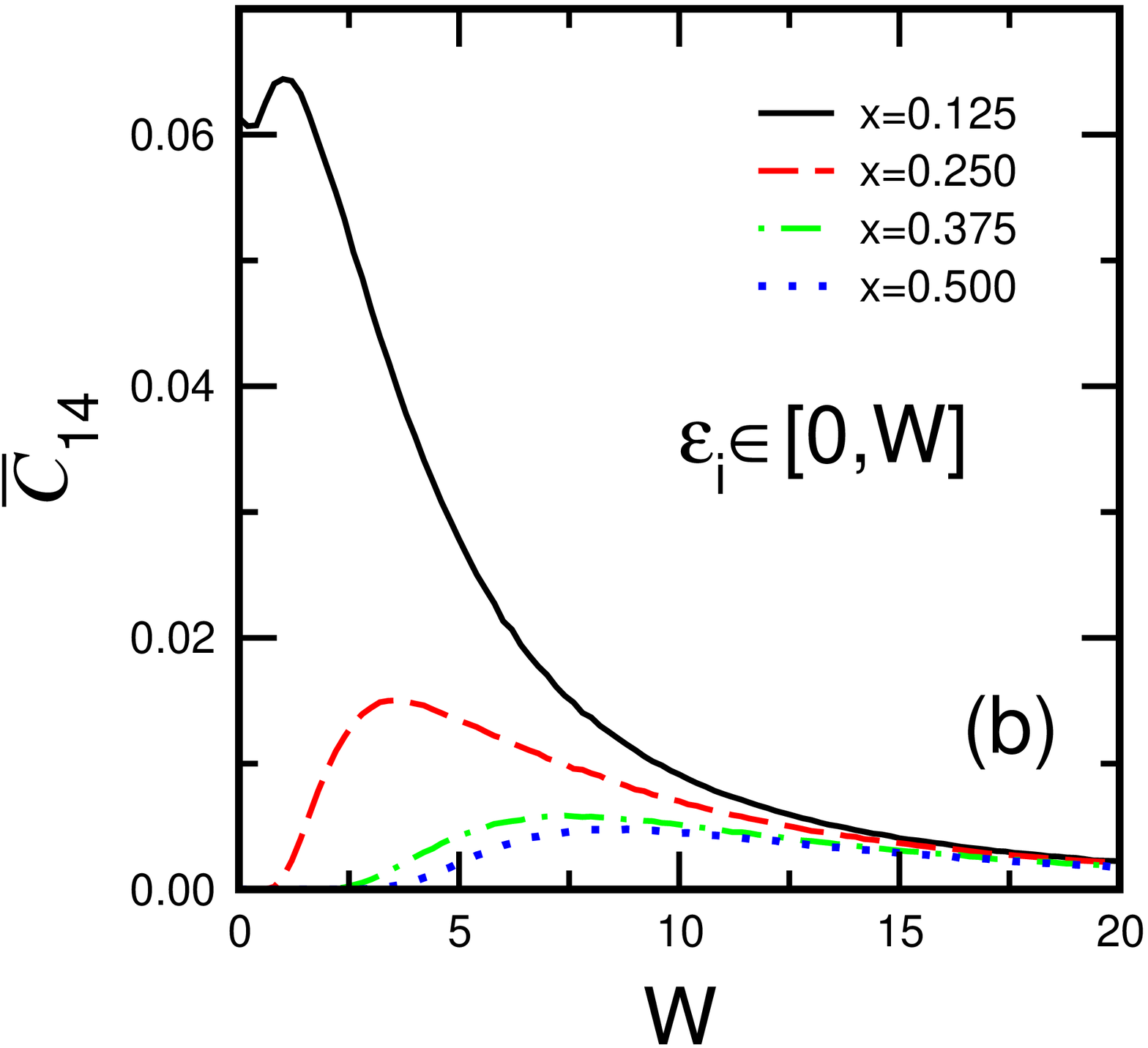} }
\caption{(Color online) 
Third neighbor average concurrence $\bar C_{14}$ of a
ring with $N_a=200$ sites as a function of band filling
$x$ for some representative values of diagonal disorder strength $W$
}
\label{fig:c14diads}
\end{figure}

\begin{figure}
\centerline{\includegraphics[scale=0.6]{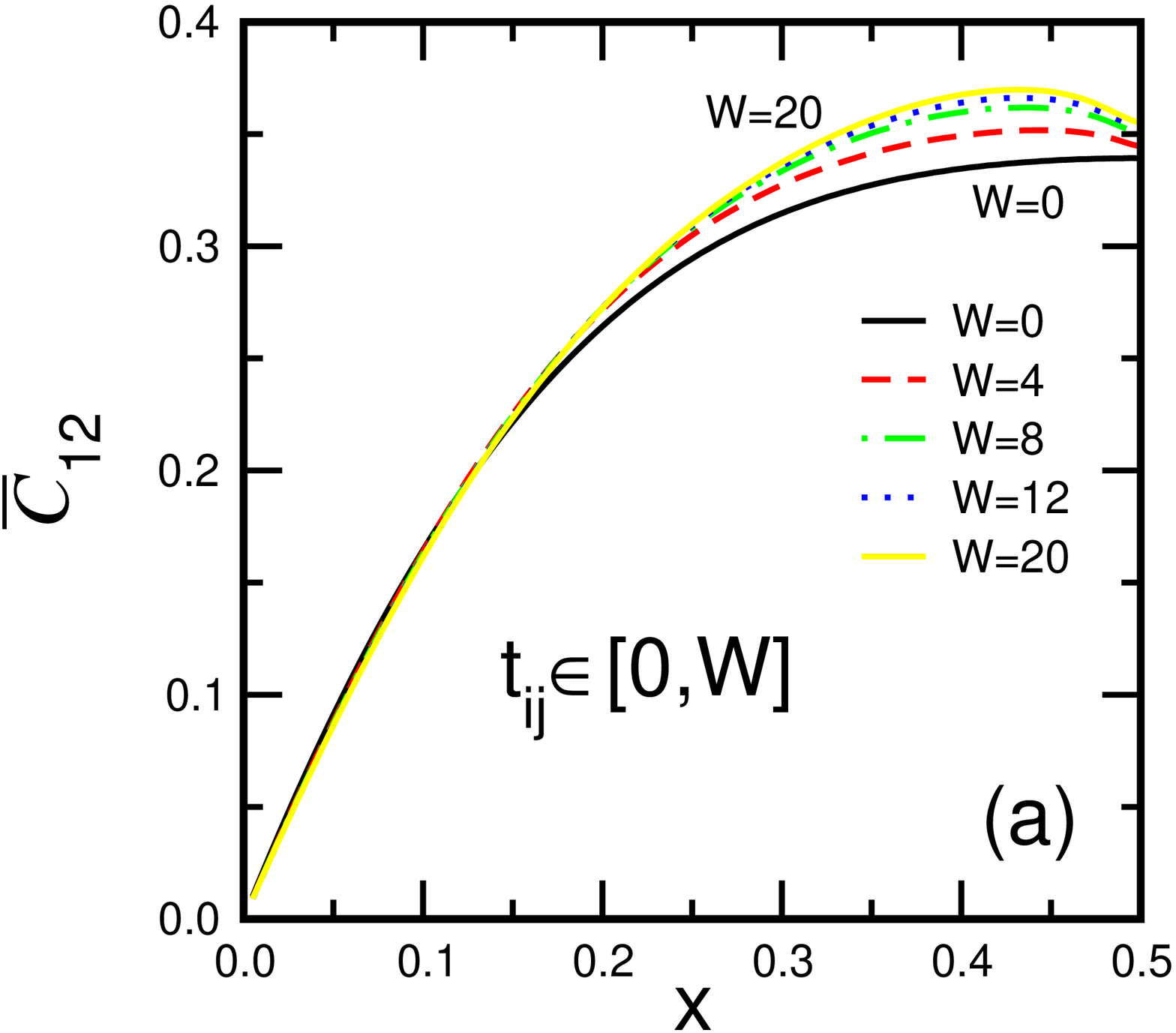}}
\centerline{\includegraphics[scale=0.6]{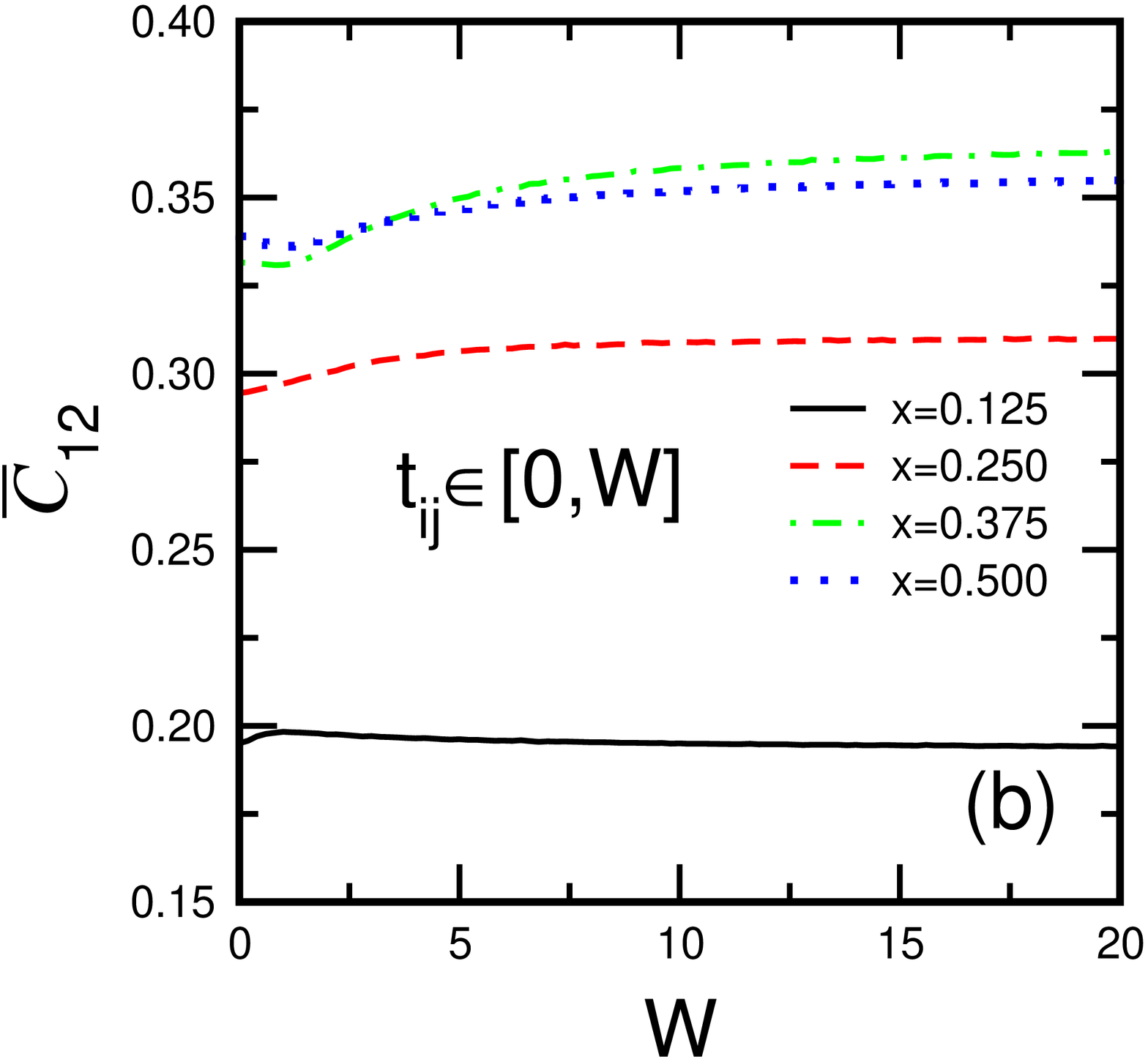} }
\caption{(Color online)  
Nearest neighbor average concurrence $\bar C_{12}$ of a
ring with $N_a=200$ sites as a function of band filling $n$
for several representative values of the off diagonal
disorder strength $W$.}
\label{fig:c12odiadis}
\end{figure}

\begin{figure}
\centerline{\includegraphics[scale=0.6]{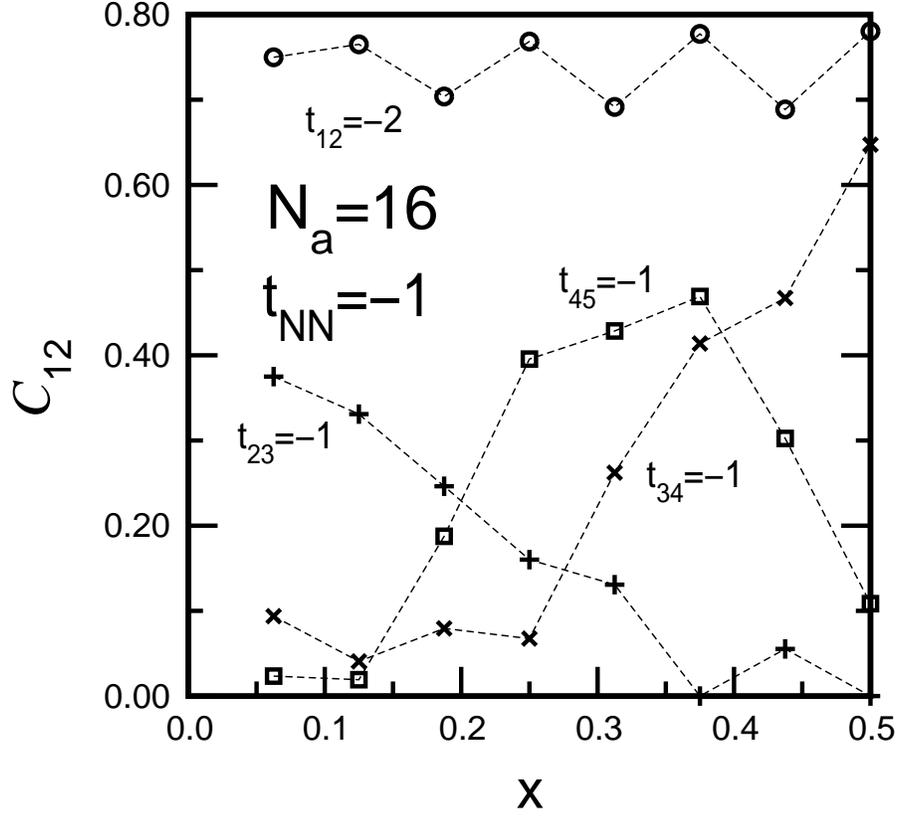} }
\caption{Nearest-neighbor concurrence $C_{12}$ for a ring with
$N_a=16$ sites and one impurity. The impurity is localized
between the sites 1 and 2 ($t_{12}=-2$). The hopping integral between
the other nearest neighbor sites is $t_{\rm NN}=-1.0$}
\label{fig:c12oneimp}
\end{figure}

\begin{figure}
\centerline{\includegraphics[scale=0.6]{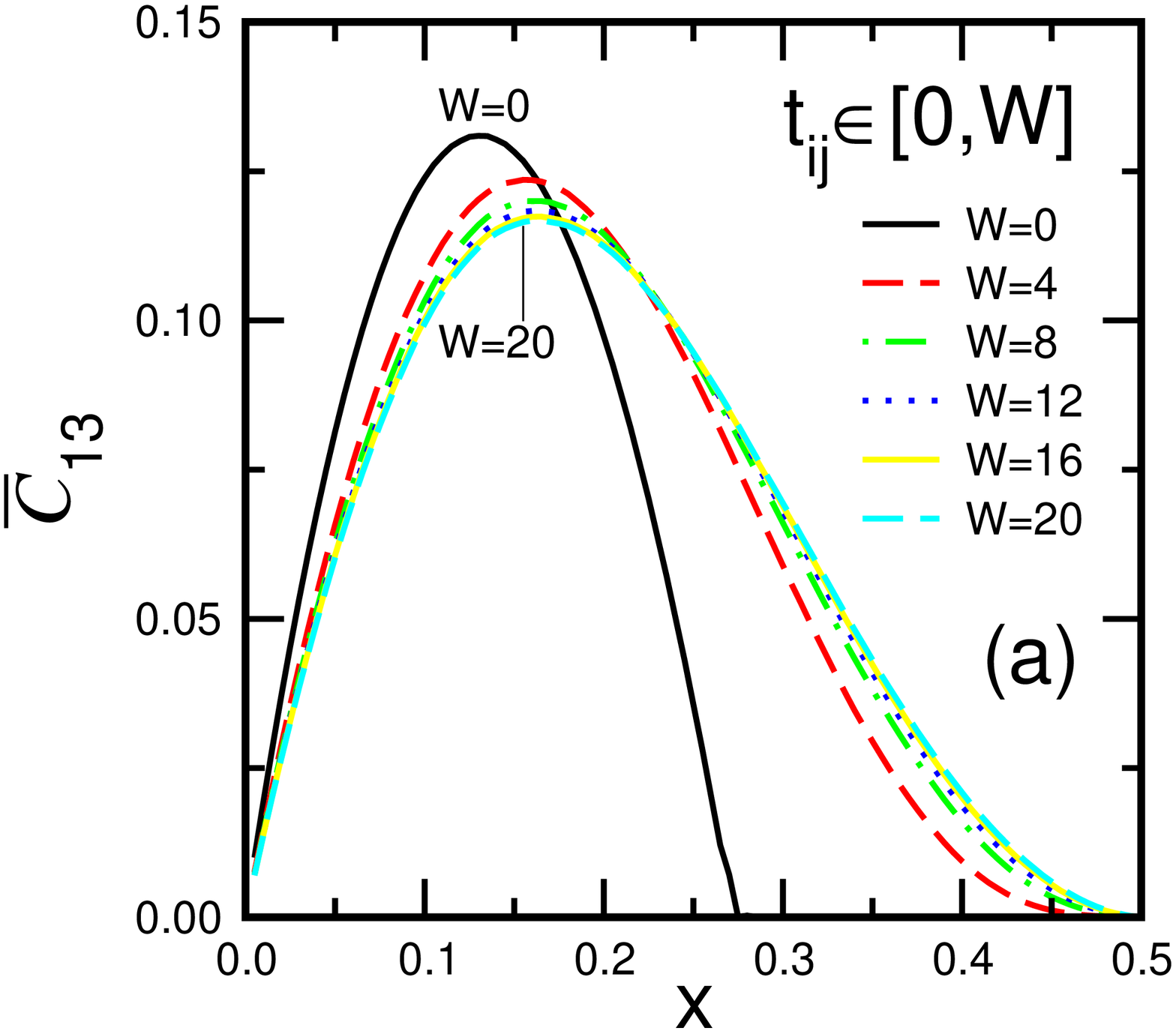}}
\centerline{\includegraphics[scale=0.6]{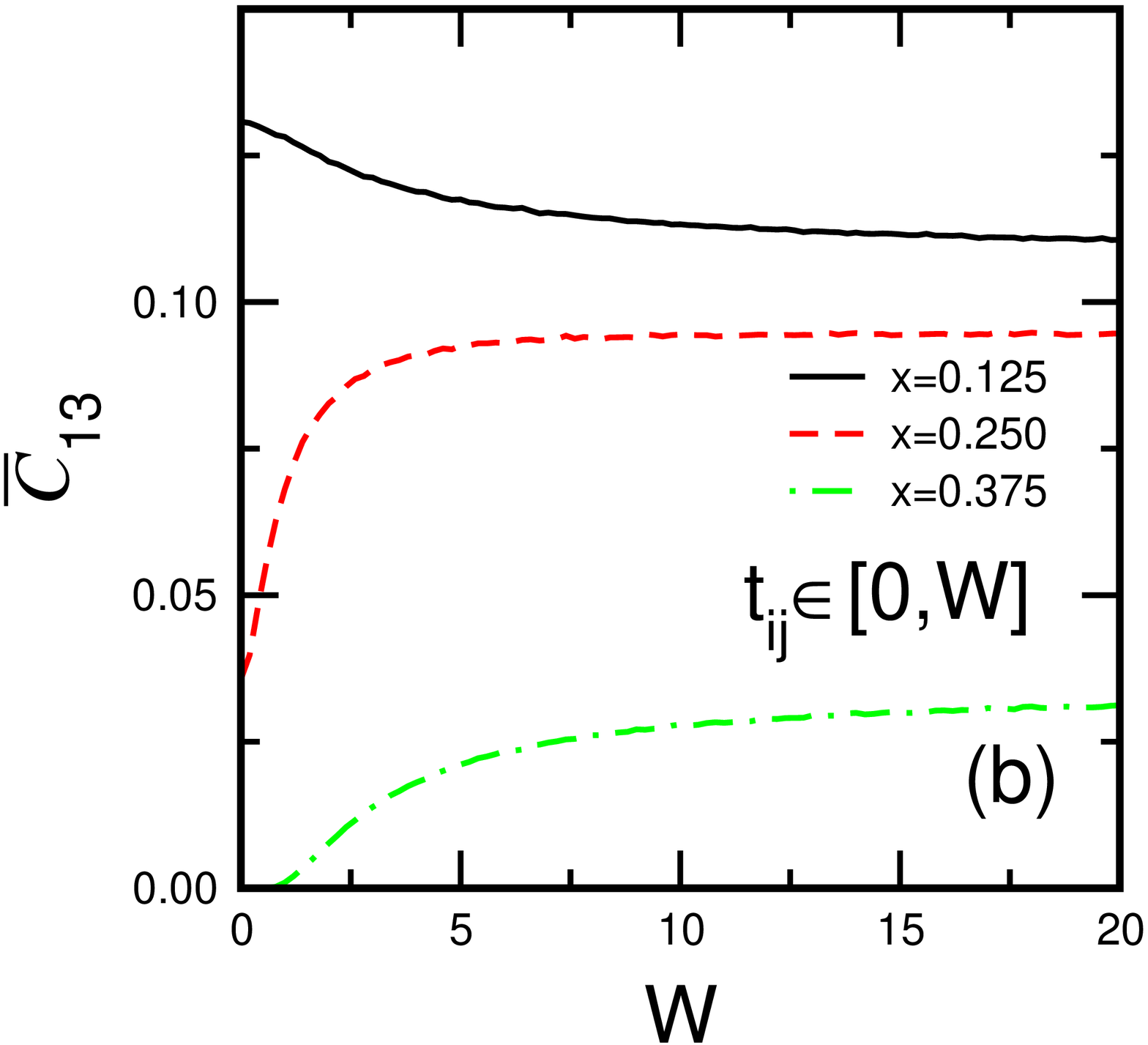} }
\caption{(Color online) 
Next nearest neighbor average concurrence $\bar C_{13}$ of a
ring with $N_a=200$ sites as a function of band filling $n$
for several representatives values of the off diagonal
disorder strength $W$.}
\label{fig:c13odiadis}
\end{figure}

\begin{figure}
\centerline{\includegraphics[scale=0.6]{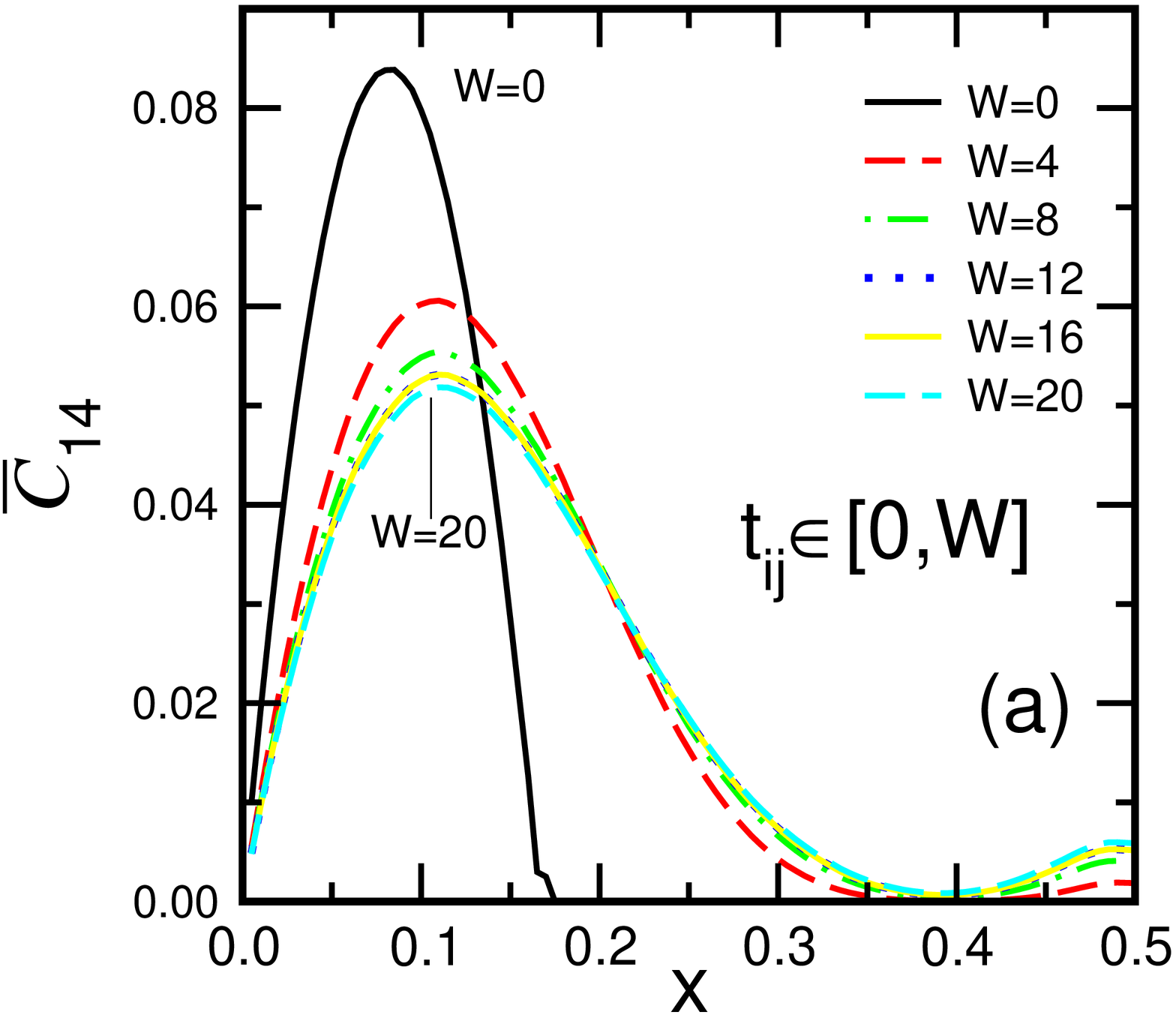}}
\centerline{\includegraphics[scale=0.6]{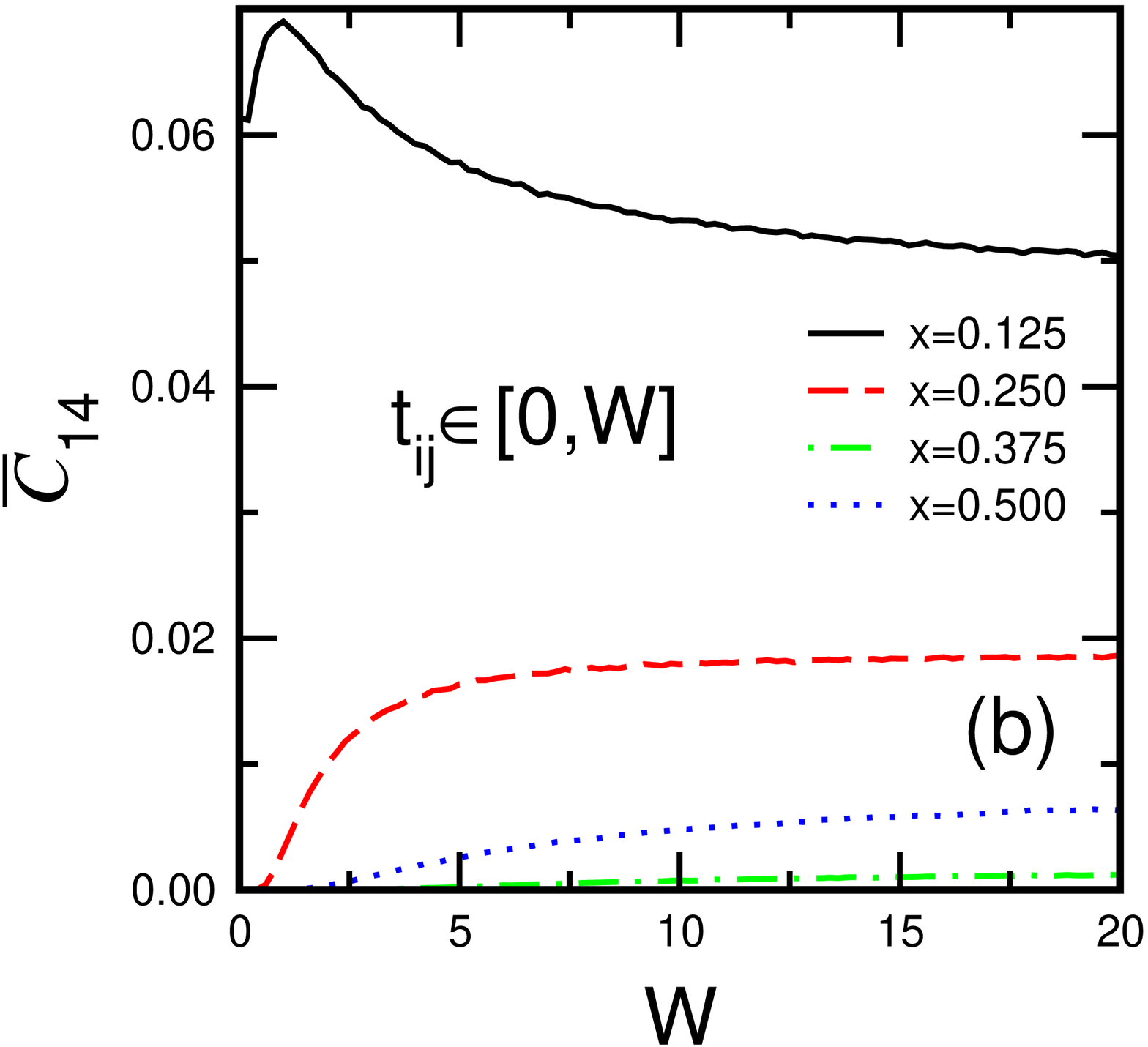} }
\caption{(Color online)  
Third neighbor average concurrence $\bar C_{14}$ of a
ring with $N_a=200$ sites as a function of band filling $n$
for several representatives values of the off diagonal
disorder strength $W$.}
\label{fig:c14odiadis}
\end{figure}

\end{document}